\documentclass[twocolumn,secnumarabic,amssymb, nobibnotes, aps, prd]{revtex4}
\usepackage{graphicx}

\begin{document}
\title{About Essence of the Wave Function on Atomic Level and in Superconductors}
\author{A.V. Nikulov}
\affiliation{Institute of Microelectronics Technology and High Purity Materials, Russian Academy of Sciences, 142432 Chernogolovka, Moscow District, RUSSIA.} 
\begin{abstract} The wave function was proposed for description of quantum phenomena on the atomic level. But now it is well known that quantum phenomena are observed not only on atomic level and the wave function is used for description of macroscopic quantum phenomena, such as superconductivity. The essence of the wave function on level elementary particles was and is the subject of heated argument among founders of quantum mechanics and other physicists. This essence seems more clear in superconductor. But impossibility of probabilistic interpretation of wave function in this case results to obvious contradiction of quantum principles with some fundamental principles of physics.
 \end{abstract}

\maketitle

\narrowtext

\section*{Introduction}

  The wave equation proposed by Erwin Schrodinger has allowed to describe numerous atomic and other phenomena but no consensus was and is about the essence of the wave function both between founders and modern experts in quantum mechanics \cite{QCh2005}. The assumption of Louis de Broglie on the matter waves seems to be corroborated by two-slit interference and other experiments but no universally recognized interpretation of these experiments was proposed for the present. The two-slit interference experiments manifest most obviously a duality in quantum physics but the essence of this duality is also the point at issue. Niels Bohr elevated the duality to the level of a principle he termed complementarity. This principle has satisfied some founders of quantum mechanics and most physical community. But other founders were sceptical. Einstein wrote to Schrodinger "It offers the believers a soft resting pillow from which they are not easily chased away" \cite{LetEin28}. Schrodinger agreed, saying that "Bohr wants to 'complement away' all difficulties" \cite{LetSchro}.

 The main point at issue was and is the problem of realism and possibility of a complete description of quantum phenomena. Einstein attacked the Copenhagen interpretation of quantum formalism since he kept firm belief in the aim of science as the discovery of the objective reality \cite{QCh2005}. He believed that the reality exists and stated that any complete theory should provide {\it the complete description of any (individual) real situation (as it supposedly exists irrespective of any act of observation or substantiation)} \cite{Einstein}. According to this position, known as {\it realism}, measured variables exist before measurement, even if they are hidden from us, i.e. we can not measure they. Contrary to this Bohr stated that all hope of attaining a unified picture of objective reality must be abandoned. This scientific debate has already long history \cite{QCh2005} and is still in progress up to now \cite{QTRF}. The works \cite{Neumann,EPR,Bohm52,Bell64,Bell66,KochSpec,Bellexp} are particular importance in history of this debate. 

 Einstein, Podolsky and Rosen \cite{EPR} revealed first non-locality of the Copenhagen interpretation of the quantum formalism. Motivated by the EPR paper \cite{EPR} Erwin Schrodinger wrote in \cite{Schr1935}: {\it Maximal knowledge of a total system does not necessarily include total knowledge of all its parts, not even when these are fully separated from each other and at the moment are not influencing each other at all}. He coined the term {\it entanglement of our knowledge} to describe this situation \cite{Zeilin01}. Entanglement, according to Schrodinger the essence of quantum
  mechanics \cite{Zeilin01}, can be easy describe in the limits of the quantum formalism by a wave function. But it is much more difficult to answer on the question: "What could this function describe?" For example, in the version of the EPR paradox, developed by David Bohm \cite{Bohm51}, the function 
$$\Psi(r_{1},r_{2}) = \psi_{\uparrow}(r_{1})\psi_{\downarrow}(r_{2})-\psi_{\downarrow}(r_{1})\psi_{\uparrow}(r_{2})$$ 
of the singlet state of two particle with spin 1/2 entangles the spin states independently of the distance $r_{2} - r_{1}$ between the particles. This version was used by Bell \cite{Bell64} for the discovery a contradiction between measurement results predicted by a realistic local theory and the quantum formalism. As it is well known the experiments made with entangled photons \cite{Bellexp,OtherExp} testify to the quantum formalism and against a realistic local theory. Nevertheless these experimental results can not answer unambiguously  on the question: {\it What does wave function describe?}

\section{What does wave function describe?}
We can say, following Bohr, that the wave function can predict only a probability of experimental result and any other questions are no sense. We can also say, following Schrodinger, that the wave function describes our knowledge. In this interpretation the EPR paradox is not paradox but we must renounce on any objective reality. The Bell's experiments \cite{Bellexp,OtherExp} leave also a possibility of a realistic interpretation of the wave function. Bell was against locality but not against realism and his favorite example of a hidden-variables theory, Bohm theory \cite{Bohm52}, is explicitly nonlocal. Also other theories were developed, for example contextual probabilistic approach to quantum mechanics \cite{Andrei}.

In contrast to the position of most physicists on the Bohr-Einstein debate J. Bell {\it felt that Einstein's intellectual superiority over Bohr, in this instance (the EPR paradox), was enormous; a vast gulf between the man who saw clearly what was needed, and the obscurantist}. Bohr stated that the atomic world differs in essence from our everyday macroscopic world and just therefore a realistic and complete description can not be valid on the atomic level. Bohr considered quantum physics as atomic physics. Therefore one ought expect that his interpretation of the wave function can come into collision with additional difficulties at description of macroscopic quantum phenomena such as superconductivity. It was noted already \cite{Legget85} that a direct extrapolation of quantum mechanics to the macroscopic level conflicts with some principles. In contrast to atomic level the wave function describes in superconductor no a probability but a density. Richard Feynman interpreted the wave function in superconductor as "classical" \cite{Feynman}. This "classical" nature of the wave function results to challenge to some fundamental principles of physics.

\subsection{What does wave function describe on the atomic level?}
Different interpretations of the wave function are most valid for different cases. For example the Bohr interpretation is very valid for description of electron orbits in atom.

 \subsubsection{Atom wave functions}
 The Schrodinger equation allowed to describe all paradoxical experimental results concerning atom orbits. It also can solved an important problem. There was a logical difficulty in the Bohr model until electron considered as a particle having a velocity since it was impossible to answer on the question: {\it What direction has this velocity?} The atom wave function does not say anything on a velocity or any distinguished direction. Einstein considered this as weakness of quantum formalism: {\it The weakness of the theory lies in the fact, that it leaves time and direction of the elementary process to chance} (the citation from the paper \cite{Zeilin9F}). But thanks to this {\it weakness} the Bohr quantization does not violate symmetry on the atomic level. The wave functions, solutions of the Schrodinger equation, give quantum numbers, the principal one $n$ giving energy of electron on atom orbits, orbital one $l$ giving the orbital angular momentum and the magnetic quantum number $m$ giving the projection of the orbital angular momentum. In fact, these quantum numbers describe all results of measurements and the wave functions give nothing in addition to these numbers having a relation to measurement results. The Bohr positivism point of view corresponds to highest degree this experimental situation. There is no sense to say not only of a velocity direction but also on a direction of the orbital angular momentum. Nevertheless the law conservation is applicable to the angular momentum. The interpretation of the magnetic quantum number $m$ as a $l$ projection along a specified axis is interpretation in terms of hidden variables. Thus, the Bohr interpretation keeps entirely symmetry in atom and this corresponds to the experimental corroboration of the conservation laws.

 \subsubsection{Interference experiments}
 The wave function $\Psi = A \exp\frac{i(pr-Et)}{\hbar}$ seems more real in the two-slit interference experiments. The probability 
$$P(y) = |\Psi|^{2} = |A_{1} \exp\frac{i(pr_{1}-Et)}{\hbar}+ A_{2} \exp\frac{i(pr_{2}-Et)}{\hbar}|^{2}$$ 
$$ = A_{1}^{2} + A_{2}^{2} + 2A_{1}A_{2}\cos(\frac{pwy}{d\hbar})$$ 
gives the interference patter predicted along $y$ on a screen placed on a distance $d$ from two slits separated of a distance $w$. In accordance with this prediction all experiments give the interference patter corresponding to the momentum $p$ of particles. The interference patter appears just as a probability when particles pass one by one through the two-slit system \cite{electro,neutron,atom}. But there is unsolved problem connected with the general unsolved problem on the collapse of the wave function at measurement. The Copenhagen interpretation refuses to answer on the question, how a particle can make its way through two slits at the same time. The Schrodinger interpretation of the wave function as a description of our knowledge is more suitable for this obvious non-locality. Some double-slit interference experiments \cite{Zeilin99} may be interpreted as a consequence  of complementarity of our knowledge. But the two-slit interference problem should be considered as unsolved. The interference experiments raise also other fundamental problems. The interference of Bose-Einstein condensates \cite{IntBEC} and atoms \cite{IntAtoms} reveals a gentle hint of challenge to the law of momentum conservation. This hint is less gentle in the Aharonov-Bohm effect \cite{Bohm59}

\subsubsection{The Aharonov-Bohm effects} The interference patter $|\Psi|^{2} = A_{1}^{2} + A_{2}^{2} + 2A_{1}A_{2}\cos(\Delta\varphi)$ depends from the difference $\Delta\varphi=\varphi_{1} -\varphi_{2}$ of the phases $\varphi_{1} = \int dt \frac{E_{1}}{\hbar} $, $\varphi_{2} = \int dt \frac{E_{2}}{\hbar} $ of the wave function $\Psi = A_{1}\exp(-i\varphi_{1}) + A_{2}\exp(-i\varphi_{2})$ describing two ways motion of a particle. Since the energy of a particle with a charge $q$ depends on electric potential $V$ and its kinetic energy is $E_{kin} = (p-qA)^{2}/2m$ the phase difference, and consequently the interference patter, can be change by a potential $V$ along one of the two ways or a magnetic flux $\Phi = \oint_{l}dl A$ inside the closed line $l$ formed by two ways of the particle from source to a screen point: $|\Psi|^{2} = A_{1}^{2} + A_{2}^{2} + 2A_{1}A_{2}\cos(\Delta\varphi_{0} + q\int dt V)$ or $|\Psi|^{2} = A_{1}^{2} + A_{2}^{2} + 2A_{1}A_{2}\cos(\Delta\varphi_{0} + q\Phi/\hbar)$. Here 
$$p = \hbar\nabla\varphi=mv + qA \eqno{(1)}$$ 
is the canonical momentum including the magnetic vector potential $A$, $\Delta\varphi_{0}$ is the phase difference at $V=0$ and $\Phi = 0$.

Thus, the Aharonov-Bohm effects arise directly from the quantum formalism. But it is no mere chance that this effect was discovered by David Bohm who introduced in 1952 a quantum potential \cite{Bohm52}. The Aharonov-Bohm experiments demonstrate that the interference pattern can be altered without any forces having acted on the particle. This experimental paradox can be easy described by the Bohm theory \cite{Bohm52}. The interpretation of the entanglement and the wave function in this hidden-variables theory is contrary to the Schrodinger one. The hidden variables in the Bohm interpretation are simply the real configuration-space coordinates of real particles, guided in their motion by a real field, the Bohm quantum potential 
$$Q_{B} = -\frac{\hbar^{2}}{4m}(\frac{\bigtriangledown ^{2}P}{P}- \frac{(\bigtriangledown P)^{2}}{P^{2}})$$
 where $P(r) = |\Psi (r)|^{2}$ is a probability field. The quantum force $F_{Q}= - \nabla Q_{B}$ influences on the particle trajectories and therefore, for example, particles passing through a two-slit system give a interference patter \cite{Hiley}. 

The Bohm interpretation of quantum mechanics, favorite example of a hidden-variables theory for John Bell \cite{Mermin93}, is {\it hideous} nonlocal. The quantum potential $Q_{B}(r_{1},r_{2},r_{3},…t) $ of several particles making on a particle, for example $r_{1}$, with the force $\bigtriangledown _{1}Q_{B}(r_{1},r_{2},r_{3},…t)$ depends on the instantaneous positions $r_{2},r_{3}....$ of all the other particles of the system described by the wave function $\Psi (r_{1},r_{2},r_{3},…t)$. Bohm has resolved the EPR paradox and the the two-slit interference problem introducing a real non-locality. The quantum non-locality in the Aharonov-Bohm effect \cite{Bohm59} is obvious \cite{Bohm61}. But few physicist can admit that the non-locality and force-free momentum transfer could be real. Some authors \cite{Wiseman} try to interpret the nonlocal force-free momentum transfer observed in the classical double-slit Aharonov-Bohm experiments and other experiments in the spirit of complementarity. But such interpretation is not valid for superconductors, when the wave function can not be considered in the spirit of Schrodinger as description of our knowledge.

\subsection{What does wave function describe in superconductors?}
All quantum phenomena observed in superconductor structures can by described by a wave function $\Psi(r) = |\Psi(r)|\exp (i\varphi (r))$ introduced by V.L. Ginzburg and L.D. Landau \cite{GL}. The phase gradient $\nabla\varphi$ is connected with the canonical momentum (1) as well as at the quantum description of elementary particles. But the $|\Psi|^{2}$ value is interpreted as a density of superconducting pairs but no as a probability. Such interpretation allows to explain the Meissner and quantization effects as a consequence of the demand that the complex pair wave function $\Psi(r)$ must be single-valued at any point of the superconductor 
$$\oint_{l}dl p = \oint_{l}dl \hbar\nabla\varphi=m\oint_{l}dl v + q\oint_{l}dl A$$ 
$$ = m\oint_{l}dl v + q\Phi = \hbar2\pi n \eqno{(2)}$$ 
The current of pairs with the charge $q = 2e$ having a density $j_{s} = 2e|\Psi|^{2}v$ shields magnetic field in superconductor depth where the pairs velocity $v = 0$. The Meissner effect $\Phi = 0$ takes place according to (2) if the wave function has no singularity inside $l$ and therefore the quantum number $n = 0$. When the wave function has singularity $n \neq 0$ and the flux quantization $\Phi = n\Phi_{0} = n\hbar \pi /e$ takes place at a strong shielding, whereas at weak shielding the velocity circulation should have  discrete permitted values $$\oint_{l}dl v = \frac{\hbar2\pi}{m} (n - \frac{\Phi}{\Phi_{0}}) \eqno{(3)}$$ 
Because of the velocity quantization (3) the persistent circular current \cite{Blatt61} $I_{p} = sj_{p} = s2e|\Psi|^{2}v$ should be observed in thin-walled ring or cylinder with a section $s$ when a magnetic flux inside it is not divisible by the flux quantum $\Phi_{0} = \hbar \pi /e$. The first experimental evidence of this quantum phenomenon was obtained as far back as 1962 \cite{LP1962}.

\section{Force-free momentum transfer in superconductor}
The persistent current $I_{p} =  s2e|\Psi|^{2}v$ is much more real than a path of a single particle through a two-slit system and the wave function in superconductor can not be interpreted as description of our knowledge. Therefore the force-free momentum transfer observed is absolutely real and there is experimental evidence of a real challenge to the law of momentum conservation. For example we see in the Little-Parks experiment \cite{LP1962} that an equilibrium direct circular current $I_{p} \neq 0$ is observed at a non-zero resistance $R > 0$ and without the Faraday voltage $d\Phi /dt = 0$. The only explanation \cite{QForce} of this paradoxical phenomenon is based on the change of the momentum circulation $\oint_l {dl p} $ of superconducting pairs on $2\pi \hbar n-2e\Phi  =2\pi \hbar (n-\Phi /\Phi _{0})$, from $2e\Phi$ to $2\pi \hbar n$, at closing of wave function in the ring with radius $r$, $l = 2\pi r$. This momentum change in a time unity $F_{q} = \hbar (n-\Phi /\Phi _{0})\omega /r$ at a reiterate switching between superconducting states with different connectivity of wave function with a frequency $\omega $, named in \cite{QForce} {\it quantum force}, replaces the Faraday voltage and restores the force balance at an average. But the real momentum transfer $\hbar (n-\Phi /\Phi _{0})/r$ because of quantization $\oint_l {dl p} = 2 \pi n $ is force-free and nonlocal. The quantum force in superconductor \cite{QForce} is more real than the Bohm quantum potential on atomic level. It has other experimental evidence \cite{Lett07} in additional to \cite{LP1962}. The problem with the force balance is also at the Meissner effect \cite{Hirsch} and the Hall effect in superconductors \cite{Berger}. Therefore a challenge to the law of momentum conservation is well deserving of attention.

\section{Force-free momentum transfer and intrinsic breach of symmetry}
The law of momentum conservation is based on symmetry demand, for example between opposite directions. Therefore it is important to note that the experimental results \cite{Dub2003} has given evidence of intrinsic breach of right-left symmetry. The quantum oscillations in magnetic field of sign and value of the dc voltage  $V_{dc}(\Phi /\Phi _{0})  \propto I_{p}(\Phi /\Phi _{0}) \propto <n>-\Phi /\Phi _{0}$ observed on segments of asymmetric superconducting rings \cite{Dub2003} have proved that the Bohr quantization breaks symmetry on the macroscopic level in contrast to the atomic level. This fundamental difference between atomic and macroscopic levels may be connected with a possibility to make a switch between states with different connectivity of wave function in superconducting loop whereas it is impossible in atom \cite{FFP8}.

\section*{Acknowledgement}
This work has been supported by a grant "Quantum bit on base of micro- and nano-structures with metal conductivity" of the Program of ITCS department of RAS and a grant of the Program "Quantum Nanostructures" of the Presidium of Russian Academy of Sciences.

\end{document}